\documentclass[11pt]{article}
\usepackage{iap2000,epsfig}

\def\be{\begin{equation}}
\def\ee{\end{equation}}
\def\bea{\begin{eqnarray}}
\def\eea{\end{eqnarray}}

\begin{document}
\title{THE X-RAY EVOLUTION OF CLUSTERS OF GALAXIES TO z=0.9\\ [16pt]}
%\mbox{\includegraphics[]{jones.eps}}}

\author{LAURENCE R. JONES$^{1}$, H. EBELING$^{2}$, C. SCHARF$^{3}$, 
E. PERLMAN$^{4}$,\\
D. HORNER$^{5}$, B.W. FAIRLEY$^{1}$, S. ELLIS$^{1}$, G. WEGNER$^{6}$, 
M. MALKAN$^{7}$}

\address{$^{1}$School of Physics \& Astronomy, University of Birmingham, Birmingham, 
UK\\
$^{2}$ Institute for Astronomy, 2680 Woodlawn Dr, Honolulu, HI
96822, USA\\
$^{3}$ Space Telescope Science Institute, Baltimore, MD 21218,
USA.\\
$^{4}$ 
Joint Ctr. for Astrophysics, Physics Dept,
University of Maryland, Baltimore County,
Baltimore, MD  21250, USA\\
$^{5}$ Lab for High Energy Astrophysics, Code 660,
NASA/GSFC, Greenbelt, MD 20771, USA.\\
$^{6}$ Dept. of Physics \& Astronomy, Dartmouth College, 
% 6127 Wilder Lab., 
Hanover, NH 03755, USA.\\
$^{7}$ Dept. of Astronomy, UCLA, Los Angeles, CA 90024, USA.\\
}

%\author{H. Ebeling,  C. Scharf, E. Perlman,
%D. Horner, B. Fairley, S. Ellis, G. Wegner, M. Malkan
%}

\maketitle\abstracts{
The  evolution of the X-ray luminosity function of clusters of galaxies
has been measured to z=0.9 using over 150 X-ray selected 
clusters discovered in the WARPS survey. 
We find no evidence for evolution 
of the luminosity function at any
luminosity or redshift. 
The observations constrain the evolution of the 
space density of moderate luminosity
clusters to be very small, and much less than predicted by most 
models of the growth of structure with $\Omega_m$=1.
All the current X-ray surveys agree on this result.
Several notable luminous clusters at z$>$0.8 have been found,
including one cluster which is more  luminous (and is probably
more massive) than the well known MS1054 cluster.
}

\section{Introduction}
The evolution of the space density of clusters of galaxies is a 
measurement sensitive
to the physical and cosmological parameters of models of structure 
formation. We describe a deep X-ray survey of clusters of galaxies 
(the Wide Angle ROSAT Pointed Survey or WARPS),
and use it to measure the evolution of the 
X-ray luminosity function (XLF) of clusters of galaxies
(Scharf et al 1997, Jones et al 1998, 
Ebeling et al 2000a, Fairley et al 2000).

\section{Survey method}
Our goal was to compile a complete, and as  unbiased as possible, X-ray 
selected sample
of clusters and groups of galaxies from serendipitous detections in deep,
high-latitude ROSAT
PSPC pointings. The spectroscopic follow-up of the clusters is now
99\% complete. The redshift range is 0.03$<$z$<$0.92 and 
half of the 157 clusters are at z$>$0.3.

A detailed review of the X-ray source detection 
algorithm used (Voronoi Tessellation and Percolation or VTP), 
the sample selection and flux correction techniques are given in 
Scharf et al (1997). The basic technique was to identify cluster candidates
via their spatial extent in X-rays, and confirm and measure redshifts
via optical imaging and spectroscopy.  We estimate a core radius for
each cluster individually in order to extrapolate the surface
brightness and derive a total flux. 
%VTP is particularly well suited to the 
%detection and characterization of low surface brightness emission
%and to the recognition of extended sources. 
At the maximum off-axis angle used (15 arcmin) 
the PSPC point-spread function degrades to 
$\approx$55 arcsec (FWHM, at 1 keV) and there is a possibility that 
some clusters at the edge of the PSPC fields (where most of the 
survey area is) are unresolved.
To reduce this possible incompleteness, our optical follow-up 
observations are not limited to extended X-ray sources but also 
include likely point X-ray sources without obvious optical counterparts.

The flux limit of the complete sample 
is 6x10$^{-14}$ erg cm$^{-2}$ s$^{-1}$ (0.5-2 keV)
and the total solid angle 73 deg$^2$. 
Detailed simulations were performed
to derive the 
survey sensitivity as a function of both source flux and source extent.

The optical imaging follow-up program is limited in wavelength
to the I band, and does not extend into the near IR. At redshifts
z$>$1.3 the I band samples the rest frame U band or bluer and K-corrections
for elliptical galaxies become very large. We expect that
this is the major reason why we have not detected clusters
at z$>$1. 

\begin{figure}[tbp]
    \centering
    \includegraphics[height=10cm]{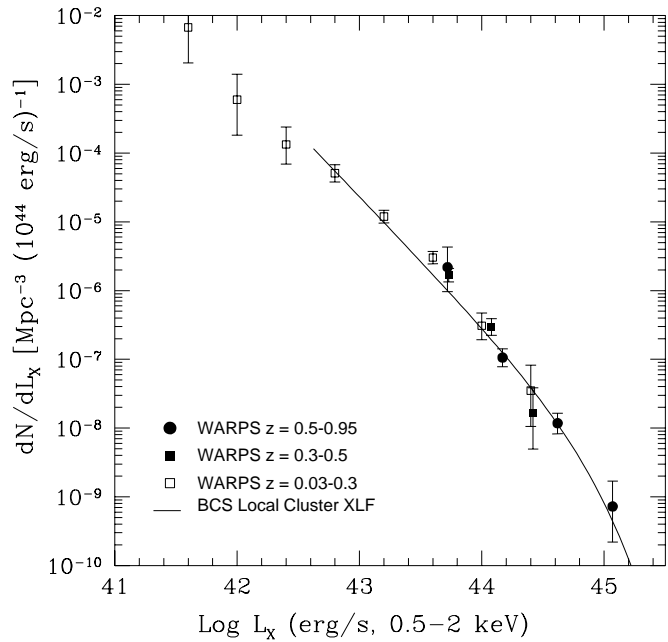}
    \caption{The X-ray luminosity function of clusters and groups 
of galaxies from the WARPS survey. No evolution of
the space density of clusters is observed up to redshifts of z=0.95. }
\end{figure}
% \begin{figure}
% \psfig{figure=xlf.ps,height=10cm}
% \caption{The X-ray luminosity function of clusters and groups 
% of galaxies from the WARPS survey. No evolution of
% the space density of clusters is observed up to redshifts of z=0.95. 
% }
% \end{figure}

\begin{figure}[tbp]
    \centering
    \includegraphics*[height=10cm]{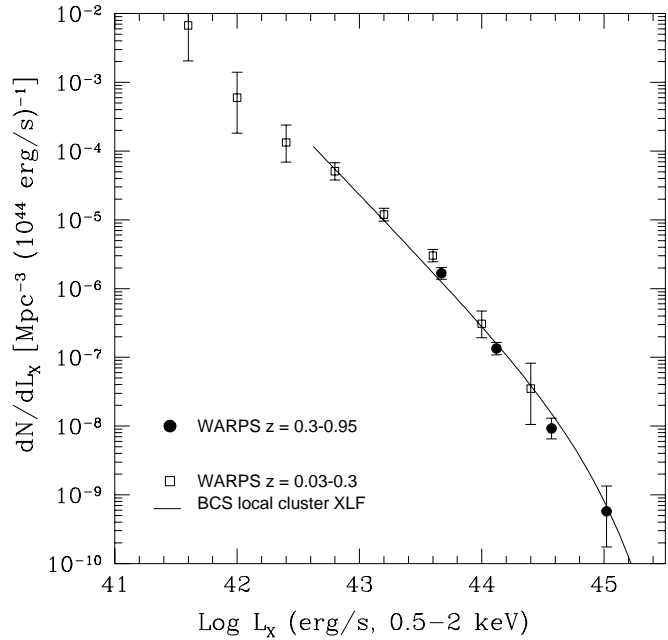}
    \caption{This figure contains the same data as figure 1, but 
combined into one redshift bin for z$>$0.3, in order to emphasize the
small statistical errors.}
\end{figure}

% \begin{figure}
% \psfig{figure=,}
% \caption{
% This figure contains the same data as figure 1, but 
% combined into one redshift bin for z$>$0.3, in order to emphasize the
% small statistical errors.
% }
% \end{figure}
% 

\section{The X-ray luminosity function of clusters}
The cluster XLF is shown in Fig 1. 
A value of q$_0$=0.5 has been assumed in calculating the luminosities
and volumes but the results are 
qualitatively similar for q$_0$=0.1.
The WARPS data points extend over more than 3 decades of luminosity.
The low redshift points are in good agreement with the BCS XLF
of Ebeling et al (1997), shown as a solid line, and extend 
the  low redshift XLF to the low luminosities of groups.

At higher redshifts, there is no evidence for evolution of the 
XLF at any redshift up to z=0.95. The error bars are based on a 
Poissonian distribution of the 
number of clusters in each bin. The survey area for each cluster has been 
calculated using the cluster flux but making the simplifying assumption
that all clusters have the same observed surface brightness profile 
(characterised as a constant angular core radius and $\beta$=2/3 index).
Mean differences from the true survey areas are small ($<$10\%).

As the XLFs at z=0.3-0.5 and z=0.5-0.95 are consistent with each other, in 
Figure 2 we
have combined both redshift ranges to increase the statistical
accuracy at the expense of a rather broad redshift bin.  Nevertheless,
the small error bars on the high redshift points (filled circles) 
indicate that any evolutionary factor $>$1.5 in the space density of moderate 
luminosity clusters (L$_X\approx$10$^{44}$ erg s$^{-1}$)  is ruled out.

The number of high luminosity (L$_X>3$x$10^{44}$ erg s$^{-1}$) clusters
in the survey at 0.7$<$z$<$1 is 5$^{+3.4}_{-2.2}$. This is 
consistent with the no-evolution 
predictions of 6.9-11, depending on whether the Ebeling et al (1997)
or de Grandi et al (1999) low redshift XLF is used. The latest
results from the REFLEX survey (Bohringer, these proceedings),
suggest that a third no-evolution prediction would be nearer the lower number
of 6.9.

The XLFs shown here are based on an initial analysis; while some details
will change (eg approximate correction for AGN contamination) in a final version,
we do not expect there to be major changes.

\section{Newly discovered luminous, high redshift clusters}
We have discovered 5 clusters at z$>$0.8, all with spectroscopically
confirmed redshifts. Four of them have X-ray luminosities similar
to, or much larger than, the Coma cluster.  One in particular,
J1226+3332, (z=0.888) has a bolometric X-ray luminosity of 
8x10$^{45}$ h$_{50}^{-2}$ erg s$^{-1}$ (q$_0$=0.5; Ebeling et al 2000b). This luminosity
is slightly higher than that of MS1054, the best known example
of a high redshift massive cluster. Our preliminary velocity dispersion
for J1226+3332, 1600 km s$^{-1}$, is also consistent with a high
mass (or the presence of substructure along the line of sight). 
The existence of more than one cluster of high mass at these redshifts 
is unlikely to be consistent with $\Omega_m=1$.

\section{Discussion}

\subsection{Comparison with results of X-ray cluster surveys}

Several recent deep X-ray clusters surveys are described in these proceedings 
and elsewhere. In the regime where these surveys have good statistical
accuracy (ie. moderate X-ray luminosities $\sim$10$^{44}$ 
erg s$^{-1}$) there is excellent agreement that {\it no evolution of the 
XLF is observed to z$\approx$0.8}. Five surveys agree on this point 
(EMSS, RDCS, SHARC, CfA, WARPS) and the only disagreement  
(the RIXOS survey of Castander et al 1995) can be understood in terms of 
the RIXOS source detection algorithm.

At the higher X-ray luminosities of the most massive clusters 
(L$_X>$5x10$^{44}$ erg s$^{-1}$), there
is some disagreement between the results of different surveys as to
the degree of evolution found at z$>$0.3 (if any). This
may partly be due to the small numbers of high luminosity clusters in any 
one survey. The range of evolution found is not large: 
from none to negative evolution 
of a factor $\approx$3. 

The WARPS and RDCS (Borgani et al, these proceedings) are both 
consistent with no evolution of the XLF up to
z=1. WARPS is not sensitive to clusters at z$>$1 because our 
optical followup does not extend into the NIR.

\subsection{Revisiting the EMSS survey}

In an effort to understand the EMSS results of Gioia et al (1990)
and Henry et al (1992), who found negative evolution at high
luminosities but relatively low redshifts (z$\approx$0.33),
we have 
remeasured the X-ray luminosities of the 11 EMSS clusters at z$>$0.3 
for which deep ROSAT
PSPC data exists, extending the work started by Jones et al (1998).
If the X-ray luminosities of EMSS z$>$0.3 clusters have been 
significantly underestimated by a factor of $\approx$2, then the 
EMSS XLF will move toward
a no-evolution result.
 
We use large (3 Mpc radius, H$_0$=50) apertures, removing point
sources and using ASCA temperatures where necessary. We find that the mean ratio 
$L_{X,PSPC}/L_{X,EMSS}\approx1.6$ and that the ratio is correlated
with the core radius measured from the PSPC images. This suggests
that the assumption of a constant core radius of 250 kpc in the EMSS 
(a reasonable assumption, given the data available at the time)
has led to an underestimate of the luminosities of clusters with
larger core radii. 

Henry (2000) has investigated EMSS cluster luminosities and
notes that the original EMSS extrapolation of the surface brightness
profile to give the total
flux (a mean correction factor of $\approx$2.5) did not take into
account the Einstein IPC psf. Including the effect of the psf
revises the EMSS luminosities upwards by a factor of 1.37, explaining
a major part of the discrepancy with the ROSAT luminosities.  However,
Henry (2000) also notes that luminosities measured with ASCA 
are only 17\% higher
than EMSS luminosities.   

%\subsection{Cosmological implications}
%Most models of the growth of structure in the Universe with $\Omega_m$=1 
%predict rapid evolution of the space density of clusters of galaxies
%(eg. scaling relation models \cite{B97} or a cold+hot dark
%matter model \cite{B94}). The tight
%constraints on the level of evolution allowed by the observations 
%rules out most of the $\Omega_m$=1 models. In models with $\Omega_m\approx$0.3
%(with or without a cosmological constant) the predicted evolution is much 
%less rapid and these values of  $\Omega_m$ are consistent with the 
%lack of evolution observed.

\subsection{Future work}
Future work will concentrate on the handful of clusters in the 
WARPS survey at z$\approx$0.8, on measurements of the evolution of the   
cluster X-ray temperature function using Chandra and XMM, and studies 
of galaxy evolution in X-ray selected clusters.  
We are also studying the WARPS sample of `fossil' groups of galaxies,
to help understand the formation of luminous elliptical galaxies.

New wide area X-ray surveys, designed to detect the rare high luminosity,
massive clusters in large numbers at high redshifts (z$>$0.5) are planned 
(see Lumb \& Jones 2000 and Ebeling et al 2000c).
It is the most massive 
clusters which have the greatest leverage to constrain cosmological
parameters.

\section*{Acknowledgments}
We acknowledge useful discussions with Pat Henry.

\end{document}